\documentclass[pra,twocolumn,showpacs,preprintnumbers,amssymb]{revtex4}

\newcommand{\one}{\mbox{$1 \hspace{-1.0mm}  {\bf l}$}}
\newcommand{\ket}[1]{\left|{#1}\right\rangle}
\newcommand{\bra}[1]{\left\langle{#1}\right|}

\newcommand{\ketbrad}[1]{\left|{#1}\rangle\!\langle{#1}\right|}

\newcommand{\modtwo}{\!\oplus\!}

\usepackage{dcolumn} 
\usepackage{amsmath,amsfonts,amssymb,graphicx}

\begin{document}


\title{On the role of memory errors in quantum repeaters}

\author{L. Hartmann$^{1}$, B. Kraus$^{1}$, H.-J. Briegel$^{1,2}$ and  W. D{\"u}r$^{1,2}$}

\affiliation{$^1$ Institut f{\"u}r Theoretische Physik, Universit{\"a}t Innsbruck,
Technikerstra{\ss}e 25, A-6020 Innsbruck, Austria\\
$^2$ Institut f\"ur Quantenoptik und Quanteninformation der \"Osterreichischen Akademie der Wissenschaften, Innsbruck, Austria.}
\date{\today}

\begin{abstract}
We investigate the influence of memory errors in the quantum repeater scheme for long-range quantum communication. We show that the communication distance is limited in standard operation mode due to memory errors resulting from unavoidable waiting times for classical signals. We show how to overcome these limitations by (i) improving local memory, and (ii) introducing two new operational modes of the quantum repeater. In both operational modes, the repeater is run blindly, i.e. without waiting for classical signals to arrive. In the first scheme, entanglement purification protocols based on one-way classical communication are used allowing to communicate over arbitrary distances. However, the error thresholds for noise in local control operations are very stringent. The second scheme makes use of entanglement purification protocols with two-way classical communication and inherits the favorable error thresholds of the repeater run in standard mode. One can increase the possible communication distance by an order of magnitude with reasonable overhead in physical resources. We outline the architecture of a quantum repeater that can possibly ensure intercontinental quantum communication.
\end{abstract}


\maketitle


\section{Introduction}
From all fields of quantum information science, quantum communication is most likely to reach a commercial application first. For long-distance communication one faces the problem that quantum channels like optical fibers are noisy and lossy, and both the output and the fidelity of the quantum information sent decrease exponentially with distance. Since quantum information can not be amplified, standard techniques from classical communication technology can not directly be used to overcome this problem. In principle, quantum error correction techniques can protect the quantum information while it is sent through a channel~\cite{Kn96}. However, the small tolerable error rates limit the length of the channel drastically before error correction must be applied. Hence, one would need a large number of segments to cover a certain distance. The requirements on the quality of measurements and local operations are also very stringent ($10^{-4}$), far below experimentally achievable accuracy today. 

Entanglement can be a resource to overcome this problem. If party $A$ holds one part of a maximally entangled pair of qubits, and party $B$ the other part quantum information can be transferred by teleportation~\cite{Be93}. When these parties are far away from each other, and channels and local operations are noisy, the problem arises how to distribute the entangled pairs among them.
 
The quantum repeater~\cite{Br98,Du98} (see also \cite{Ch05,Kl06,exp1,Exp,Exp2,Exp3,Du06R}) is a solution to this problem based on entanglement purification~\cite{Be96,De96,Pa01,Kw01,Pa04,Ya03,Re05,Re06} and entanglement swapping~\cite{Be93,Zuxx}. The distance $L$ between the parties $A$ and $B$ is divided into smaller segments such that one can send parts of maximally entangled pairs through the channel that do emerge with sufficiently high fidelity for entanglement purification. Noisy local operations and measurements do not allow to purify one single maximally entangled pair from several copies, but the fidelity can be increased for remarkably high errors in the local operations and measurements on the order of percent~\cite{Du98}. Via entanglement swapping, segments are connected, establishing entangled pairs over larger distances. Observe that the connection process again decreases the fidelity such that one may connect only a few segments before the entanglement can no longer be increased by purification. The key ingredient of the quantum repeater is to use the combination of purification and entanglement swapping in a nested scheme, i.e. on different repeater levels. After few connections are made, the resulting pair is again purified by several copies obtained in the same way. Then the sequence ``connection/re-purification" is repeated until one has reached the desired distance between the parties.
Most importantly, the physical and temporal resources needed for the quantum repeater scale only polynomially with the distance between parties $A$ and $B$. The details naturally depend on the errors, the specific purification protocol, and the repeater meta-protocol, i.e. the distribution and number of repeater stations and their individual setup. The repeater protocols range from the standard protocol ~\cite{Be96,De96}, where all pairs needed in the process are created initially as an ensemble (maximal physical, minimal temporal resources), over the ``Innsbruck protocol"~\cite{Du98} (physical resources scale logarithmically with the distance) to the ``Harvard protocol"~\cite{Ch05} with minimal physical resources (two qubits per repeater station) but maximal temporal resources. For practical purposes minimal physical resources are desirable since it is hard to control or even establish a large number of interacting quantum systems. In this light, one would tend to prefer the last two of the protocols above.

While in previous investigations the influence of noise in channels and in local control operations has been extensively studied, memory errors have not been included in the analysis so far. It was implicitly assumed that (almost perfect) local memory is available by some means. If this assumption is valid, as can e.g. be ensured by using local encoding to actively maintain quantum information, one obtains a scalable scheme that allows for quantum communication over arbitrary distances with polynomial overhead. However, all repeater schemes require the storage of pairs before they are further processed, and the influence of imperfect memory needs to be studied. 
In particular, at high repeater levels when long-distance pairs are processed, the waiting times can be significant. Estimated times to establish an entangled pair over, say, intercontinental distances are of the order of the decoherence times of the best known memory systems today, making the consideration of memory errors a necessity. In this paper we address the problem of memory errors in quantum repeaters. Specifically we demonstrate 
\begin{itemize}
\item[(i)] the limits of the quantum repeater with memory errors when run in standard mode (error detection mode), where we show that memory errors lead to a limited communication distance.
\item[(ii)] ways to reduce or overcome memory errors by using decoherence free subspaces or local encoding for storage.
\item[(iii)] a novel operational mode for the quantum repeater, the error correction mode, which in principle allows one to overcome the limitations of memory errors, however suffers from low error thresholds.
\item[(iv)] a blind operation mode and hybrid architectures that allow one to increase the possible communication distance by an order of magnitude, without changing favorable error thresholds. 
\end{itemize}

The paper is organized as follows. In the next section we briefly describe the building blocks of a quantum repeater, entanglement purification and swapping. We sketch different repeater protocols and present the error model we will use. In section~\ref{limits} we apply the error model, especially memory errors, to the quantum repeater. We derive the limits for communication distance when no memory-enhancing techniques are used and discuss error thresholds. As a possible way to overcome the limitations due to memory errors, a direct solution is to reduce or even eliminate them, which we discuss in  Sec.~\ref{reduceMemory}. If no perfect quantum memory is available, we show in Sec.~\ref{blind} that blind mode is an alternative way to relax the limitations of the quantum repeater. We then outline possible architectures for a quantum repeater in section~\ref{architecture}, and summarize our results in Sec.~\ref{summary}.

\section{Basic principles}
\label{basicprinciples}
We start with some notations, present purification protocols, entanglement swapping and repeater protocols for both a repeater in error detection as well as in error correction mode~\cite{As03}, and introduce the error model we are going to use.
\subsection{Notation}\label{notation}
Throughout the paper we will speak of two spatially separated parties $A$ and $B$, who share certain entangled pairs of qubits between them. We denote these pairs by $A_1B_1,...,A_NB_N$, i.e. $A$ holds the qubits $A_1,...,A_N$, while $B$ holds $B_1,...,B_N$. Whenever it is not clear form the
context on which system an operator is acting, we specify it
with a sub- or superscript. An operation is called local if it acts only on $A$'s or only on $B$'s qubits, e.g. $U_{\rm CNOT}^{A_1\to A_2}$ is a local \textsc{cnot}-operation with qubit $A_1$ as control and $A_2$ as target~\cite{Dur_noteCNOT}. By $P_{\Phi}$ we denote a projector
onto the states $\ket{\Phi}$. Furthermore, $\sigma_i$ denote the
Pauli operators, explicitly:
$\sigma_0=\one,\sigma_1=\sigma_x,\sigma_2=\sigma_y,
\sigma_3=\sigma_z$. The Bell states are denoted by
$\ket{\Phi_j}=\one\otimes \sigma_j \ket{\Phi^+}$ with
$\ket{\Phi^+}=1/\sqrt{2}(\ket{00}+\ket{11})$.

Instead of the usual Bell states we often take their graph state equivalents~\cite{He06}, which we call graph Bell states.
The graph state basis for two qubits defined in the basis $\ket{0}_z$, $\ket{1}_z$ (eigenbasis of the Pauli $\sigma_z$-operator) and in the basis $\ket{0}_x$, $\ket{1}_x$ (eigenbasis of the Pauli $\sigma_x$-operator) is
\begin{align}\label{graphbasis}
\ket{00}_G&:=2^{-1/2}(\ket{00}_{zx}+\ket{11}_{zx})\nonumber\\
\ket{01}_G&:=2^{-1/2}(\ket{01}_{zx}+\ket{10}_{zx})\nonumber\\
\ket{10}_G&:=2^{-1/2}(\ket{00}_{zx}-\ket{11}_{zx})\nonumber\\
\ket{11}_G&:=2^{-1/2}(\ket{01}_{zx}-\ket{10}_{zx}).\nonumber
\end{align}
Expressions like $\ket{00}_{zx}$ mean $\ket{0}_z\otimes\ket{0}_x$. The graph state basis is related to the standard Bell basis $|k_1,k_2\rangle_B$ by a Hadamard operation in $B$. When the basis is clear from the context we will omit the label $G$. If such a state is for example the first pair shared between $A$ and $B$ we write $\ket{00}_G^{A_1B_1}$. 

We will consider density matrices that are diagonal in the graph state basis, 
\begin{equation*}
\rho=\sum_{k_1,k_2=0}^1\lambda_{k_1,k_2}\ketbrad{k_1,k_2},
\end{equation*}
and we will sometimes write $\rho=\sum_{k_1,k_2=0}^1\lambda_{k_1,k_2}P_{k_1,k_2}$ with a projector
\begin{equation*}
P_{k_1,k_2}:=\ketbrad{k_1,k_2}.
\end{equation*}

We denote by $(m_1,m_2)$ a possible shift of the basis, i.e. a permutation of the basis vectors. That is,
\begin{equation*}
\rho=\sum_{k_1,k_2} \lambda_{k_1,k_2} \ketbrad{k_1\modtwo m_1,k_2\modtwo m_2},
\end{equation*}
where $\oplus$ will always mean addition modulo $2$. 
We remark that, without loss of generality, any density matrix can be brought to a graph-diagonal form without changing the diagonal coefficients by applying appropriate sequences of (probabilistic) local operations. To be precise, these operations correspond to the stabilizing operators of the given graph, in our case $K_1,K_2,K_1K_2, \one$ with $K_1=\sigma_x\otimes\sigma_z, K_2=\sigma_z\otimes\sigma_x$~\cite{DurLectureNotes}. Permutations of basis vectors can be achieved by local unitary operations of the form $\sigma_z^{m_1} \sigma_z^{m_2}$. 
Note that the state $\rho$ results from sending one part of a graph state $\ket{k_1,k_2}$ through a Pauli-diagonal channel
\begin{equation*}
{\cal E}_1(\rho)=\sum_{i=0}^3 p_i \sigma_i\rho\sigma_i,
\end{equation*}
with $p_0=\lambda_{00}$, $p_1=\lambda_{10}$, $p_2=\lambda_{11}$, and $p_3=\lambda_{01}$.

Later, we will use the Werner states \cite{We89}
\begin{equation}\label{wernerstate}
\begin{split}
\rho_W(x)&:= x \ket{00}_G\!\bra{00} + (1-x)/4 \one\\
&:=F\ket{00}_G\!\bra{00}+(1-F)/3\sum_{i,j\neq 0,0}\ket{ij}_G\!\bra{ij}
\end{split}
\end{equation}
with $F=(3x+1)/4$, which are uniquely defined by the quantity $F$, the fidelity, whereas more general graph diagonal states are usually only fully specified by all diagonal coefficients. We call the largest of these the fidelity, and we will often omit the other coefficients in the discussion. This simplification is justified since the purification protocol we will use produces states close to particular graph diagonal states, so called binary mixtures $\lambda_{00}\ket{00}_G\!\bra{00}+\lambda_{10}\ket{10}_G\!\bra{10}$. Here, $\lambda_{10}=1-\lambda_{00}$ such that binary mixtures are also specified by only one coefficient.


\subsection{Entanglement purification}

Entanglement purification allows one to produce from several noisy copies of an entangled state a few copies with high fidelity by means of local operations and classical communication. For perfect operations, the fidelity can, in principle, be brought arbitrarily close to unity. However, many purification steps are required for nearly perfect pairs, so that, in practice, only some finite fidelity is achievable (``finite" meaning smaller than one). If the local operations required in the purification process are noisy themselves, then even in principle no perfect pairs can be obtained. At this stage, what matters to us is that in practice no protocol will produce perfect, maximally entangled pairs. Besides the maximal fidelity we can reach, there is also some minimal fidelity we need for the purification process. This minimal fidelity depends on the protocol we use for the purification, and it is called the purification threshold. 

A number of different protocols exist, which differ in their purification range (i.e. the set of states they can purify), the efficiency, and the number of copies of the states they operate on~\cite{DurLectureNotes}. We present two-way entanglement purification, i.e. a purification protocol using two-way classical communication, namely the \textsc{dejmps}-protocol~\cite{De96}, and also one-way entanglement purification based on Calderbank-Shor-Steane codes.

\subsubsection{Two-way entanglement purification}

We take a recurrence protocol for purification, where we consider the \textsc{dejmps}-protocol~\cite{De96} since it has a very good efficiency in terms of convergence speed and robustness. Remarkably, the fidelity of states can be significantly increased even if errors in operations and measurements are on the order of percent. For the moment, however, we consider perfect operations and measurements, and generalize the formulae later when we will have introduced our error model. The protocol operates on two entangled pairs, and can be viewed as a generalization of the recurrence entanglement purification protocol introduced in Ref.~\cite{Be96}. We slightly modify the protocol as compared to the original work such that it purifies graph diagonal Bell states rather than Bell states. This corresponds, however, to a simple change of local basis which does not modify the protocol as such.
The protocol consists of the following steps.\\

\noindent (i) depolarization of the density matrix to graph diagonal form; in fact this step need not be executed since off-diagonal elements do not influence the change in the diagonal elements and converge to zero upon iteration of the protocol.\\  

\noindent (ii) local basis change $|0\rangle_z \to \frac{1}{\sqrt{2}} (|0\rangle_z -i|1\rangle_z), |1\rangle_z \to \frac{1}{\sqrt{2}} (|1\rangle_z -i|0\rangle_z)$ in $A$ and $|0\rangle_x \to \frac{1}{\sqrt{2}} (|0\rangle_x +i|1\rangle_x), |1\rangle_x \to \frac{1}{\sqrt{2}} (|1\rangle_x +i|0\rangle_x)$ in $B$. The effect of this basis change on two graph Bell states is, omitting an irrelevant phase factor, 
\begin{equation*}
\begin{split}
&\ket{x_1,x_2}\ket{y_1,y_2}\rightarrow\ket{x_1,x_1\modtwo x_2}\ket{y_1,y_1\modtwo y_2}.
\end{split}
\end{equation*} 

\noindent (iii) application of bilateral local \textsc{cnot}-operations $U_{\rm CNOT}^{A_1 \to A_2} \otimes U_{\rm CNOT}^{B_2 \to B_1}$, such that 
\begin{equation*}
\begin{split}
&\ket{x_1,x_2}\ket{y_1,y_2}\rightarrow\ket{x_1\modtwo y_1}\ket{x_2,y_1,x_2\modtwo y_2}.
\end{split}
\end{equation*}

\noindent (iv) local measurement of qubit $A_2$ [$B_2$] in the eigenbasis of $\sigma_z$ [$\sigma_x$] with corresponding result $(-1)^{\zeta_2}$ [$(-1)^{\xi_2}$], where $\zeta_2,\xi_2 \in\{0,1\}$.\\ 

\noindent (v) decision: keep the state $\rho_{A_1B_1}$ if the measurement results indicate a successful purification round. This decision requires two-way classical communication between the parties $A$ and $B$.\\

We let the protocol act on the tensor product of two graph diagonal states $\rho_{A_1B_1}$, $\rho_{A_2B_2}$ with coefficients $\lambda_{k_1,k_2}$ and $\mu_{j_1,j_2}$ respectively, which have bases shifted by $(m_1,m_2)$, and $(n_1,n_2)$ respectively, i.e. on
\begin{equation}\label{twopairs}
\rho=\!\!\!\!\!\sum_{k_1,k_2,j_1,j_2=0}^1\!\!\!\!\!\lambda_{k_1,k_2}\mu_{j_1,j_2}P_{k_1\oplus m_1,k_2\oplus m_2,j_1\oplus n_1,j_2\oplus n_2}.
\end{equation} 
After steps $(i)$--$(iv)$, qubits $A_1$ and $B_1$ will be in the state
\begin{equation*}
\begin{split}
\rho^\prime=\!\!\!\!\!\!\sum_{k_1,k_2,j_1,j_2=0}^1\!\!\!\!\!\!\!&\lambda_{k_1,k_2}\mu_{j_1,j_2}\delta_{\zeta_2\oplus\xi_2,k_1\oplus k_2\oplus j_1\oplus j_2\oplus m_1\oplus m_2\oplus n_1\oplus n_2}\\
&\times P_{k_1\oplus j_1\oplus m_1\oplus n_1,k_1\oplus k_2\oplus m_1\oplus m_2}
\end{split}
\end{equation*}
where $\delta$ is the Kronecker-delta.
The condition for a successful purification step relates the measurement outcomes 
$\zeta_2$, $\xi_2$ and the basis shifts in the following way: $\zeta_2\oplus\xi_2=m_1\oplus m_2\oplus n_1\oplus n_2$. In case this condition is fulfilled, we arrive at a simple expression for $(\rho^\prime)_{i_1,i_2}=:\lambda^\prime_{i_1,i_2}$, namely
\begin{equation}
\label{Purymap}
\lambda_{i_1\oplus m_1 \oplus n_1,i_2\oplus m_1 \oplus m_2}^\prime=\frac{1}{N} \sum_{k_1=0}^1 \lambda_{k_1,k_1\oplus i_2} \mu_{k_1\oplus i_1,k_1\oplus i_1\oplus i_2},
\end{equation}
where $N=\sum_{i_1,i_2}\lambda^\prime=(\lambda_{00}+\lambda_{11})(\mu_{00}+\mu_{11})+(\lambda_{01}+\lambda_{10})(\mu_{01}+\mu_{10})$ is a normalization constant that quantifies the probability to obtain the corresponding measurement results. The normalization is independent of the basis shifts. 
While the basis shifts do not play a role in the present discussion of the \textsc{dejmps}-protocol, they will become crucial when running the repeater in a blind operational mode, sec.~\ref{blind}. 

The \textsc{dejmps}-map, after a successful step, always drives the states closer to a binary mixture like  $\lambda_{00}\ketbrad{00}_G+\lambda_{10}\ketbrad{10}_G$. The map is also most effective on binary mixtures, and least effective on Werner states $\rho(x)= x \ketbrad{00}_G + (1-x)/4 \one$.

There are two distinct purification strategies for which we can use the \textsc{dejmps}-protocol, regular entanglement purification and entanglement pumping. \\ 

\noindent {\bf (a)} Regular entanglement purification\\

First, we could imagine to have an ensemble consisting of several copies of some elementary, noisy pair of qubits. Whenever we perform a successful purification step on two such pairs, the resulting pair of higher fidelity goes to the next purification round, otherwise it is discarded. In the \textsc{dejmps}-map we have in this case $\lambda_{ik}^{(n)}=\mu_{ik}^{(n)}$  in every round $n$, and the (attractive) fixed point of the map is a perfect graph Bell state.
In practice, we can not do infinitely many steps to reach this fixed point, let alone that errors are present that prevent one to approach this fixed point even in principle. We call this purification strategy ``regular entanglement purification". The drawback of this strategy are the many qubit pairs we need to prepare and keep ready-to-use during the process. The number of pairs is exponentially growing with the number of purification steps we wish to perform.\\ 

\noindent {\bf (b)} Entanglement pumping\\

Second, we can always use identical, elementary pairs in each round to further purify the pair we obtained from a previous successful step. If at any time we are not successful, the whole protocol must be restarted with two fresh elementary pairs. This strategy is called entanglement pumping~\cite{Du98}. The advantage clearly is that the physical resources (qubit pairs to be stored simultaneously) stay constant. We need not count elementary pairs because they do not have to be stored but are consumed at once. The elementary pairs can rather be re-created on demand. With entanglement pumping, we have $\lambda_{ik}^{(n)}\neq\mu_{ik}^{(n)}$, except in the first round, and the $\mu_{ik}^{(n)}$ are the same in every round $n$ in the \textsc{dejmps}-map~(\ref{Purymap}). Even infinite iteration will not lead to maximally entangled pairs, but in practice (with errors in the operations), the fixed point of the map can even be closer to a maximally entangled pair than for the regular entanglement purification~\cite{Du98}. Because one saves physical resources at the expense of only a polynomial overhead in time, entanglement pumping was favored in the most recent designs of quantum repeaters~\cite{Ch05,Kl06}. The real drawback of using entanglement pumping in the quantum repeater shows up when we later include memory errors, where an -- albeit polynomial -- overhead in time becomes a problem.

We remark that this is also the reason why we do not consider nested entanglement pumping~\cite{Du2003}. Nested entanglement pumping has the same fixed point of the purification map as regular entanglement purification. The number of pairs grows only linearly with the nesting level at the expense of a temporal overhead exponential in the number of purification steps one performs on each nesting level. Although the fixed point is (nearly) reached for about $3$ nesting levels, the additional temporal overhead make this purification scheme unfavorable in the presence of memory errors.\\

\subsubsection{One-way entanglement purification}\label{OnewayEPP}

In his PhD thesis \cite{As03}, Aschauer introduced a general scheme to construct entanglement purification protocols from quantum error correction codes. In particular, for each Calderbank-Shor-Steane (CSS) code that uses $n$ physical qubits to protect $k$ qubits, one can construct an entanglement purification protocol that operates on $n$ initial copies of two-qubit states and produces $k$ purified pairs as output. As described in \cite{As03}, the purification protocols can either be run (i) in error correction mode, or (ii) in error detection mode. In case of (i), output pairs are kept deterministically and measurements on remaining pairs are used to determine the required error correction operation. This operation mode only requires one-way classical communication. For (ii), the information gathered in the measurement of $(n-k)$ pairs is used to decide whether the remaining pairs should be kept or discarded. The ones that are kept have a higher fidelity than before. This operational mode is the standard mode for recurrence protocols as discussed above. Here, we will concentrate on (i), entanglement purification run in the error correction mode.

In the following we briefly review the work by Aschauer \cite{As03}. We consider the situation where the sender, Alice, wants to send quantum information to the receiver, Bob. To this aim, Alice might either send a system, $A_0$, prepared in an arbitrary state $\ket{\Psi}$ to Bob or she might prepare a maximally entangled state between two systems, send one to Bob and use the other to teleport the state $\ket{\Psi}$ to Bob. To protect the quantum information from the errors that occur during the transmission process, quantum error correction is used in the first and entanglement purification in the second scenario.

In a quantum error correction protocol (we consider here the case where the state of a single qubit is protected) Alice prepares $n$ auxiliary systems (denoted by $A$) in a state $\ket{a_1,\ldots a_n}_A$, with $a_i\in \{0,1\}$. Then she applies the encoding operation $U_{A,A_0}$ to $A$ and the system $A_0$, prepared in the state $\ket{\Psi}$ and carrying the quantum information, and sends all systems to Bob. In the simplest case, where no errors occur during the transmission, Bob receives the systems in the state $U_{B,B_0}\ket{a_1,\ldots a_n}_B\ket{\Psi}_{B_0}$. He applies $U_{B,B_0}^{-1}=U_{B,B_0}^\dagger$ to decode the quantum information and measures the auxiliary systems in the computational basis. Finally, he will be left with a system in the state $\ket{\Psi}$.

Let us now consider an entanglement-based version of this protocol. We make use of the fact that $U_A\otimes \one_B \ket{\Phi^+}_{AB}=\one_A\otimes U_B^T \ket{\Phi^+}_{AB}$ for any operator $U$. The idea is that Alice prepares Bob's system at a distance using an entangled state. Suppose that Alice and Bob share $n+1$ maximally entangled states, $\ket{\Phi^+}_{AB}^{\otimes n}\ket{\Phi^+}_{A_0B_0}$, where $A$ ($B$) denotes the first $n$ systems of Alice (Bob) respectively, and $A_0$ ($B_0$) denotes the $(n+1)$-th system of Alice (Bob). Alice applies $U_{A,A_0}^T$ and teleports the state $\ket{\Psi}$ to Bob with the help of the $(n+1)$-th pair. It is straightforward to verify that the remaining system is then described by the state $U_{B,B_0}\ket{\Phi^+}_{A,B}^{\otimes n}\sigma_j\ket{\Psi}_{B_0}$, where $j$ depends on Alice`s measurement outcome. Thus, if Alice measures her auxiliary systems in the computational basis and tells Bob the value of $j$, Bob can apply $\sigma_j^{B_0}$ to be left with exactly the same state as in the quantum error correction model.

In order to include the errors that occur during the transmission we describe the channel by the map ${\cal E}_1$ with ${\cal E}_1(\rho)=\sum_{i=0}^3 p_i \sigma^i \rho \sigma^i$ where $\sum_i p_i=1, p_i\geq0$ (see section~\ref{notation}). We investigate the case, where all the errors occur independently on each of the sent qubits. Thus, the map we consider is ${\cal E}={\cal E}_1^{\otimes n}=\sum_{\bf i} p_{\bf i} \sigma^{\bf i} \rho \sigma^{\bf i}$, where ${\bf i}=(i_1,\ldots i_n)$, with $i_j\in \{0,\ldots3\}$ and $p_{\bf i}=p_{i_1}\cdots p_{i_n}$. In the first scenario the encoded message is sent through this channel. Receiving the systems, Bob applies $U^\dagger$ and measures the auxiliary systems. Alice sends Bob the classical information about $\{a_i\}$ which allows Bob to determine the error syndrome with which he can correct the error. In the second scenario one qubit of each maximally entangled state is sent through the channel. Then the pairs are purified to one pair which is highly entangled. This pair is then used by Alice to teleport the state $\ket{\psi}$ to Bob. Considering the purification of the image of the map, ${\cal E}$, i.e. $U_{\cal E} \ket{\psi}=\sum_{\mathbf i} \sqrt{p_{\mathbf i}}\sigma^{\mathbf i} \ket{\psi}\ket{{\mathbf i}}_R$, such that $\mathrm{tr}_R(P_{U_{\cal E} \ket{\psi}})={\cal E}(P_\psi)$, with some auxiliary system $R$, it is straightforward to show that applying entanglement purification and then teleportation is equivalent to quantum error correction, where the message is sent through the same channel. The minimal required fidelity for this entanglement purification protocol, the purification threshold, turns out to be more stringent than for two-way classical communication \cite{Be96,As03} ($F \gtrsim 0.8$ as compared to $F > 0.5$ for a protocol using two-way classical communication). However, the advantage of error correction protocols is that they are deterministic. Note that the $1$-way purification protocols in~\cite{As03} are based on the Bell $\ket{\Phi^+}$-state. One could easily make them consistent with our graph basis by applying local basis changes.

\subsection{Entanglement swapping}

\noindent Entanglement swapping \cite{Zuxx} is the operation on two maximally entangled qubit pairs, where a Bell measurement is performed on one qubit of each pair with the result that the remaining two qubits are afterwards maximally entangled. If the maximally entangled pairs are the graph Bell states $A_1B_1$ and $B_2C_1$, a Bell measurement on the qubits $B_1$, $B_2$ is e.g. realized e.g. by a \textsc{cnot}-operation $U_{\rm CNOT}^{B_1\to B_2}$ followed by $\sigma_z$-measurements on qubits $B_1$, $B_2$ with outcomes $\zeta_{B1}$, $\zeta_{B2}$, leaving $A_1$, $C_1$ in the desired maximally entangled state up to a local basis change that depends on the measurement outcomes. We remark that classical communication is required to perform a proper adjustment of the local basis at the final state.
Entanglement swapping can be viewed as a teleportation of the state of qubit $B_1$ to $C_1$. If we assume that qubit $C_1$ is at some distance from $A_1$ and $B_1$, $B_2$ are somewhere in the middle, we will often call this swapping process a ``connection" or a ``link" because the goal of the quantum repeater is to establish entanglement over larger distances, here between parties $A$ and $C$. 

If both pairs are not maximally entangled, the teleportation will be that of an imperfect pair by imperfect means, resulting in a decreased or even vanishing entanglement of the final pair between $A$ and $C$. We call this an imperfect connection or imperfect link, and it is easy to understand that the fidelity of a pair after $L$ imperfect connections is decreasing exponentially with $L$. To see this, consider non-maximally entangled pairs of Werner form, eq.~(\ref{wernerstate}).
Connecting two such pairs by means of a Bell measurement as outlined above results in a state that is diagonal in the graph state basis, and has a reduced fidelity. After depolarization of the resulting state and performing the required basis change depending on the measurement outcome, one obtains again a Werner state $\rho_W(x')$ with $x'=x^2$, i.e. the fidelity $F' = (3x'+1)/4$ is reduced quadratically. The connection of $L$ pairs yields $x'=x^L$, i.e. an exponential decrease with $L$.  

If we consider two graph diagonal pairs of the form Eq. (\ref{twopairs}), the resulting pair after the Bell measurement has coefficients 
\begin{equation}
\label{connectionmap}
\lambda_{i_1\oplus m_1\oplus n_1\oplus\zeta_{B1},i_2\oplus m_2\oplus n_2\oplus\zeta_{B2}}^\prime=\!\!\sum_{k_1,k_2=0}^1\!\!\lambda_{k_1\oplus i_1,k_2\oplus i_2}\mu_{k_1,k_2}, 
\end{equation}
\noindent where $\zeta_{B1},\zeta_{B2}$ denote the outcomes of the Bell measurements leading to a permutation of the output vector (which could be undone by performing appropriate local unitary operations of the form $\sigma_z^{\zeta_{B1}}\sigma_z^{\zeta_{B2}}$). Again, the resulting state is graph diagonal, but the basis is shifted by $(m_1\modtwo n_1\modtwo\zeta_{B1},m_2\modtwo n_2\modtwo\zeta_{B2})$, an expression that depends on the initial basis shifts and the measurement outcomes. As in the purification protocol, these random basis shifts do not matter because one simply can keep track of them without the need to actually correct them. In fact, the same sequences of operations (i.e. the same protocol for entanglement swapping) can be applied, only the basis of the resulting density matrix changes.   

The scaling of the fidelity with the number of simultaneous links becomes even worse with imperfect operations, which we have not considered yet. We will describe the map resulting from imperfect connections later after introducing our error model. For the moment we have seen that even with perfect local operations we could only connect a few pairs before the entanglement would vanish. This is where the quantum repeater comes into play, whose repeater protocol determines where to interrupt the connection process and to re-purify the involved states. We turn to repeater protocols in the following.

\subsection{Repeater protocols}
\label{repeater}

The repeater protocol governs which purification protocol to use (e.g. \textsc{dejmps}), which purification strategy (regular; pumping), and which ``geometry". By geometry we mean where to place repeater stations and with which resources to equip them depending on the purification protocol, the purification strategy, and the linking strategy, i.e. how many stations to link after one purification round is complete. 
We will describe some repeater protocols with $2$-way purification protocols that have been developed to demonstrate functionality of the quantum repeater (and which are not optimized for any specific physical implementation). \\

\subsubsection{Standard repeater protocol}
The original repeater protocol~\cite{Be96,De96} uses regular entanglement purification where all required pairs are stored in parallel and the number of purification steps on each level is constant, say $M$. The total distance is divided into $N=2^n$ segments, and after each purification round two segments will be connected such that we have $n$ repeater levels. The time for the completion of the whole repeater process is $M(2^{n+1}-1)$ in units of the time we need for the first purification step, and we have neglected gate operation times and the times we need for connections. While the total time is already determined by the standard repeater protocol, the physical resources depend on the initial, elementary pairs, the purification protocol, and the errors. In this scheme the physical resources are very demanding since all pairs ever used in the process are created right at the beginning and the required resources (i.e. total number of pairs) are given by $R=(M+1)^{n}$. Despite the fact the the required resources (i.e. parallel channels or pairs to be stored) grow only polynomially with the distance, since $R$ can be rewritten as $R = N^{\log_{2}M+1}$, the overhead can be substantial. \\

\subsubsection{Innsbruck protocol}
The Innsbruck protocol~\cite{Du98} is based on entanglement pumping using the \textsc{dejmps}-purification protocol. As in the standard repeater protocol the total distance is divided into $N=2^n$ segments. On the lowest repeater level, elementary pairs are purified, and once they have reached some sufficiently high ``working" fidelity, always two adjacent pairs are connected throughout the chain. The resulting pairs of lower fidelity must be stored, so every second repeater station needs an extra qubit for storage. On the lowest level the process of purification/connection is repeated and the resulting low fidelity pair is used to purify the one that is stored. Iteration leads to a high fidelity pair over twice the initial distance. The whole scheme is repeated on higher and higher repeater levels, and we need again extra storage qubits on every $4^{th}$, $8^{th}$ etc. repeater station. The physical resources hence grow logarithmically with the distance. Compared with the standard repeater protocol, the physical resources have been drastically reduced at the expense of a polynomial overhead in time~\cite{Du98}. Purification now takes place sequentially, where new elementary pairs at each repeater level need to be re-created using the same physical resources, and one hence needs to wait until the new elementary pair arrives. In addition, a failure in the purification process on any repeater level means that the pair in question has to be discarded, and the stochastic process to rebuild it must be started again from the lowest level. Note that this means extra waiting times for pairs on higher repeater levels that depend on the supply of pairs from the level where the failure occurred. As pointed out above, these waiting times become significant when we include memory errors.

\subsubsection{Harvard protocol}
From a practical point of view it is desirable to use the minimum of physical resources since many qubits are hard to control and to store. In that respect the Harvard protocol~\cite{Ch05}, a variant of the Innsbruck protocol, is the most advanced since it uses the minimum possible number of two qubits at each repeater station. This reduction of physical resources compared to the Innsbruck protocol is possible because the capacities of some repeater stations were not fully used in the Innsbruck protocol, but are now fully activated by an ingenious setup. We will not describe this setup here in detail, but merely note that the price for minimal resources is: (a) connection of up to $5$ pairs at once (among them $3$ elementary ones), (b) even longer waiting times for high-level qubits in case of failure. Point (a) implies that we need tighter error thresholds because otherwise $5$ connections may lead to a fidelity below the purification threshold. From point (b) follows that the limits of the Innsbruck protocol, which we are going to derive when we include memory errors, also hold for the Harvard protocol.     

\subsubsection{Protocols using purification by error correction}
In principle the above protocols could also use entanglement purification by error correction. But the purification range determined in~\cite{As03} is already small for protocols run in a concatenated way, which is the equivalent of regular entanglement purification in the error detection mode. An equivalent to entanglement pumping was not discussed, but the purification ranges would certainly be very small if not vanishing. Memory errors would thus render both approaches useless very soon. Later, we will show that we can get rid of the problem with memory errors for the case of a concatenated, error correction type purification. Hence, we will only consider the equivalent of the standard repeater protocol later.

\subsection{Error model and purification and connection with imperfect means}

\subsubsection{Error model}
We conclude the section by presenting the error model we are going to use in the rest of the paper. We emphasize that the results we obtain and in particular the conclusions we draw are independent of the details of the error model, but are rather a consequence of unavoidable waiting times when using the quantum repeater in one of its standard operational modes. What may however differ slightly are the actual numbers, where the white noise model we assume turns out to provide a rather conservative estimate of the noise threshold, in particular when compared to situations where one particular kind of noise (e.g. phase noise) is dominant and much better performance and error thresholds can be obtained.
We model imperfect operations on two qubits $x_1$ and $x_2$ as a mixture of perfect operations and white noise:
\begin{equation}\label{errormodel}
O_{x_1,x_2}(\rho)=p\, O_{x_1,x_2}^{ideal}(\rho)+(1-p)\textstyle\frac{1}{4} \one_{x_1,x_2}\otimes\mathrm{tr}_{x_1,x_2}(\rho),
\end{equation}
where $O_{x_1,x_2}^{ideal}$, the ideal two-qubit operation, has probability $p$ and the two-qubit white noise has probability $1-p$.
The measurements are based on imperfect projections described by positive operator valued measure elements $P_0=\eta \ketbrad{0}+(1-\eta)\ketbrad{1}$ and $P_1=\eta \ketbrad{1}+(1-\eta)\ketbrad{0}$. 

Finally, we use local depolarizing channels to describe memory errors, i.e. local white noise. On a single qubit the depolarizing CP-map reads
\begin{equation}\label{depmap}
(D\rho)(t)=q(t)\rho+(1-q(t))/4\sum_{k_1=0}^3\sigma_{k_1}\rho\sigma_{k_1},
\end{equation}
with $q(t)=e^{-\kappa t}$ and $\kappa$ is the inverse decoherence time. On a graph-diagonal, two-qubit density matrix $\rho=\sum_{k_1,k_2=0}^1\lambda_{k_1,k_2}\ketbrad{k_1,k_2}_G$ the map is
\begin{equation}\label{memoryerror}
[(D^{[1]}\otimes D^{[2]})\rho](t)=\!\!\sum_{k_1,k_2=0}^1\!(q^2\lambda_{k_1,k_2}+(1-q^2)/4)P_{k_1,k_2}.
\end{equation}
Now, we re-derive the \textsc{dejmps}-map and entanglement swapping for imperfect operations and measurements of the above form.

\subsubsection{Purification with imperfect operations and measurements}

When we include the errors in operations and measurements, the \textsc{dejmps}-map, equation~\ref{Purymap}, is modified. Intuitively it is clear that the errors in the measurements, $\eta$, will mix the results of a successful step with those of an unsuccessful step, while the errors in the operations, $p$, will introduce white noise. The modified formula is
\begin{widetext}
\begin{equation}\label{imperfectPurymap}
\begin{split}
\lambda_{i_1\oplus m_1 \oplus n_1,i_2\oplus m_1 \oplus m_2}^\prime={\textstyle\frac{1}{N}}\left({\textstyle\frac{1-p^2}{8}}+p^2\sum_{a=0}^1\left(\eta^2+(1-\eta)^2\right)^{a\oplus 1\oplus\zeta_2\oplus\xi_2}\left(2\eta(1-\eta)\right)^{a\oplus\zeta_2\oplus\xi_2} \sum_{k_1=0}^1 \lambda_{k_1,k_1\oplus i_2}\,\mu_{k_1\oplus i_1,k_1\oplus i_1\oplus i_2\oplus a}\right).
\end{split}
\end{equation}
\end{widetext}
\noindent Again, $\zeta_2$, $\xi_2$ are the measurement outcomes of step (iv). The normalization $N=\sum_{i_1,i_2}\lambda^\prime$ represents the probability for a successful purification step, where the criterion for success, $\zeta_2\oplus\xi_2=m_1\oplus m_2\oplus n_1\oplus n_2$, also remains the same.

As before, initial basis shifts of the two pairs simply lead to a different basis shift on the resulting pair. This fact remained true because we can still commute the local basis shifts through the Clifford operations and the Pauli errors. In this sense, local basis shifts still only lead to a re-interpretation of what successful measurement outcomes are.

\subsubsection{Entanglement swapping with imperfect operations and measurements}

So far we have concentrated on entanglement purification. The second part of the repeater protocols is the linking of farther apart stations when stations in between perform (imperfect) entanglement swapping on two pairs of graph diagonal states. With the error model from above, we expect that the measurement errors lead to an admixture of the results of the other measurement outcomes, and that the errors in the operations lead to an admixture of white noise. The modified version of equation~(\ref{connectionmap}) is 
\begin{widetext}
\begin{equation}\label{imperfectconnectionmap}
\lambda_{i_1\oplus m_1\oplus n_1\oplus\zeta_{B1},i_2\oplus m_2\oplus n_2\oplus\zeta_{B2}}^\prime=\frac{1-p}{4}+p\sum_{a,b=0}^1\left(\eta^2\right)^{(a\vee b)\oplus 1}\left(\eta(1-\eta)\right)^{a\oplus b}\left((1-\eta)^2\right)^{a\wedge b}\!\!\sum_{k_1,k_2=0}^1\!\!\lambda_{k_1\oplus i_1\oplus a,k_2\oplus i_2\oplus b}\,\mu_{k_1,k_2}
\end{equation}
\end{widetext}
\noindent where $\zeta_{B1}$, $\zeta_{B2}$ are still the outcomes of the Bell measurement, and $\vee$ is the logical \textsc{or}, $\wedge$ the logical \textsc{and}. Note again that initial basis shifts of the pairs merely result in a different basis shift of the linked pair, where the shift is now randomized by the measurement outcomes.

\section{Limits of the quantum repeater}\label{limits}

In this section we show how uncorrected errors in memory limit the maximal distance over which entangled pairs can be created. First, we study the repeater in standard mode, then in error correction mode.

\subsection{Limits of the quantum repeater in standard mode}\label{limitsEDM}

As mentioned, the standard scheme for the quantum repeater uses two-way classical communication to reveal whether purification steps have been successful or not, and only in the first case the resulting pair is kept for further processing. Otherwise, the process must be started anew. The classical signal needs time to cover the distance between the repeater stations, and this time increases on higher repeater levels, where the stations are further apart. On higher repeater levels the signal time dominates by far all other timescales such as the gate operation time. During the time needed for the classical communication, the quantum systems have to be kept in some quantum memory where they are subject to memory errors. If this quantum memory is not perfect, there is a distance between parties $A$ and $B$ that can not be exceeded in the standard quantum repeater scheme because during the time the classical signals need to cover this distance the fidelity of the entangled pairs drops below the purification threshold. Naturally, this maximal distance depends on memory errors, but also on the errors and the repeater protocol, where now protocols needing less {\em temporal} resources are favored. 

In previous work, repeater protocols were developed in a kind of ``bottom-up" strategy. With chosen error models (except memory errors) and purification protocols one created a certain base module that ensured the functionality of purification and entanglement swapping, and made sure that this module could be repeated on higher levels with polynomial scaling of time and physical resources. One can keep this point of view when one includes strategies to reduce or eliminate memory errors. This, we will discuss in Sec.~\ref{reduceMemory}. 
On the other hand, when memory errors are present, then the maximal distance is a constraint and it is more natural to adopt a ``top-down" approach. Given a distance between the parties $A$ and $B$ the question is, can we reach it and what resources does it cost us? 

Our goal in this section is to determine the maximal distance that different repeater protocols can achieve. As a first step, we look at the purification range of the \textsc{dejmps} protocol on different repeater levels. We will assume throughout that the distance between two repeater stations is $10$ km, such that a classical signal needs $0.333 \times 10^{-4}$ s to travel. Further, each higher repeater level doubles this distance and hence also the signal time. We include all errors presented in the last section into the analysis of the purification range. In a second step, we simulate the full quantum repeater, where we concentrate on the standard and Innsbruck protocol having in mind that the Harvard protocol can not perform better than the Innsbruck protocol in terms of thresholds and reachable distance.

\subsubsection{Limits of \textsc{dejmps} purification protocol on different repeater levels}\label{purVSreplevel}

In the standard schemes we must wait for the classical signals to cover the distance between the repeater stations in question before we can do the next purification step.

We want to determine the purification range of the \textsc{dejmps}-map, equation~(\ref{imperfectPurymap}), on different repeater levels when memory errors are present. The purification range lies between a lower fixed point of this map~\cite{footnote}, which we call the purification threshold, and some upper fixed point.  
  
The purification range of this map is hard to determine analytically. For fixed parameters, a numerical analysis is straightforward, and can be used to analyze the performance of the protocol and in particular the influence of memory errors. Note that we are not considering the whole repeater in the following, but isolated repeater levels. To determine the purification range on some level we iterate the map several times (strictly speaking one would need an infinite number of times). Between each application of the map we let the involved states decohere for a certain amount of time. We also choose some initial state, and the purification threshold depends on that state. For regular entanglement purification, the upper fixed point of the map is independent of the initial state, while for entanglement pumping it strongly depends on the initial state.

Here, we do a general treatment of the quantum repeater, and hence we do not use parameters of any specific, physical set-up. Since we would like to obtain tolerable errors for local operations/measurements on the order of percent we choose $p=\eta=0.99$. As coherence time we assume $\kappa^{-1}=1$ s. The coherence time has a strong influence on the purification range and even more on the whole quantum repeater, and we will demonstrate this fact in the discussion of the repeater. With repeater stations that are about $10$ km apart, such that the signal time on repeater level $1$ is $t_0=0.333 \times 10^{-4}$ s, the waiting time for a signal on the $n^{th}$ level is $2^{n-1}\times t_0$ since we assume that each level doubles the distance. The memory error, equation~(\ref{memoryerror}), will hence act for at least a time $2^{n-1}\times t_0$ on the $n^{th}$ repeater level between every purification step. We neglect gate operation times that, on higher levels, are dominated by the classical signal times. To test the purification ranges of the \textsc{dejmps}-map on different repeater levels we are going to use this minimal waiting time. 

As initial states for the \textsc{dejmps}-protocol we take Werner states $\rho_W(x)$, eq.~(\ref{wernerstate}), on each repeater level. We make this choice here and in the rest of the section, because we want to stay consistent with our error model, i.e., we also assume the channels through which we establish pairs to be subjected to white noise processes. Usually this is not true, e.g. in optical fibers we find a dominance of dephasing noise, but it is the worst choice we can make for the \textsc{dejmps}-protocol, so we are definitely not being over-optimistic. Note that any noise model for channels can be brought to white noise form without changing the channel fidelity~\cite{Du05}. 

In  Table~\ref{purificationregime} we give purification regimes for different repeater levels. The second column lists the purification threshold for regular entanglement pumping. The third column gives the maximal reachable fidelity in this case, whereas in the fourth column we give the maximal reachable fidelity using entanglement pumping with initial Werner states of fidelity $0.8$. Naturally the data will vary if one puts the actual parameters of some physical setup, but there will always be some maximal distance, which, with the chosen parameters, lies between repeater level $11$ and $12$, corresponding to about $10$-$20$ thousand kilometers between the most remote stations. That the maximal distance corresponds to these repeater levels is intuitively clear, since the signal time on the $12^{th}$ level is $2^{12}*t_0\approx0.14$s which approaches the order of the decoherence time $\kappa^{-1}=1$s. The maximal distance will go down drastically for a repeater using the Innsbruck protocol (or other qubit-saving but time-consuming) protocols. But this distance will also go down for the standard repeater protocol when there are only a finite number of purification steps and imperfect links between repeater stations. 

\begin{table}[!ht]
\begin{tabular}{c|cc|c}
rep. level&min. fidelity&max. fidelity&max. fid. (pumping)\\ \hline
1&0.5276&0.985870&0.882761\\
2&0.5276&0.985778&0.882689\\
3&0.5278&0.985595&0.882545\\
4&0.5280&0.985227&0.882257\\
5&0.5284&0.984491&0.881682\\
6&0.5292&0.983017&0.875948\\
7&0.5310&0.980056&0.878236\\
8&0.5344&0.974090&0.873666\\
9&0.5417&0.961958&0.864609\\
10&0.5575&0.936728&0.846823\\
11&0.5965&0.880294&0.812544\\
12&-&-&-\\
\end{tabular}
\caption{\label{purificationregime}Purification regimes. The first column displays the repeater level where we assume a doubling of distance with each level. The second column contains the lowest possible fidelities of Werner states that can still be purified and the third column contains the fidelity to which they can be purified. The last column shows the maximal achievable fidelities of states that are purified by entanglement pumping with Werner states of fidelity $0.8$.}
\end{table}

When we relate these results to the whole quantum repeater we realize the following:\\ 
(a) The standard repeater protocol uses regular entanglement purification, but only a few steps on each level as opposed to the infinitely many steps we apply to determine the purification range. Hence, there will be a dependence on the initial, lowest level state. But this dependence is weak and becomes less and less significant on higher levels, where more and more purification steps have been executed. Since the upper fixed point of the purification map for regular entanglement purification is independent of the initial state it translates into a general upper bound for the maximal reachable fidelity of any repeater run in error detection mode -- with the exception of blind operation, see section~\ref{blind}.\\ 
(b) Repeater protocols based on entanglement pumping, e.g. the Innsbruck protocol, start with some initial state on the lowest level, and, again, the dependence on that initial state becomes weaker on higher levels. Note however, that in the repeater process the \textsc{dejmps}-map drives the states closer to binary mixtures, on which it afterwards operates more efficiently. That is, higher repeater levels get states close to binary mixtures as their initial pumping states. The situation can be completely different when we determine the fixed points of the purification map and always use the same initial pumping state that is far from a binary mixture and closer to Werner states. Hence, these fixed points do not say much about the repeater, but they still illustrate the influence of the memory errors in a simple way.

\subsubsection{Maximal distance of different repeater protocols}

Now we have assembled all tools to analyze the quantum repeater operated in error detection mode with different repeater protocols. We do not simulate the repeater, but use the success probabilities of the purification steps to estimate the physical or temporal resources we need. In this way we obtain {\em average} values for the performance of the repeater and do not explore the worst cases when the purification on some level fails unusually many times.

For the standard repeater protocol where all pairs are initially prepared and then processed in parallel we expect to get a maximal distance close to the one where purification is no longer possible (see table~\ref{purificationregime}). On the one hand, there is the advantage that purified pairs from lower levels are already closer to a binary mixture such that the purification threshold is better than for Werner states. On the other hand, the imperfect linking of pairs is additionally decreasing the fidelity. With the same choices for the parameters as above, and executing $3$ purification steps on each level, we obtain table~\ref{levelsAndResources} showing the repeater levels, the resources (qubit pairs) needed, and the maximal fidelity we reach. The resources are easy to compute. Let $p_i^{[l]}$ be the probability to succeed in the $i^{th}$ purification step on the $l^{th}$ repeater level. These probabilities correspond to the normalization factor in the \textsc{dejmps}-map, Eq.~(\ref{imperfectPurymap}). On average we need $2/p_i^{[l]}$ pairs to get one purified pair for round $i+1$. For $3$ steps, we need $2^3/\prod_{i=1}^3 p_i^{[l]}$ pairs on level $l$, and for the whole repeater with $n$ levels we need $2^{3n}/(\prod_{l=1}^n\prod_{i=1}^3 p_i^{[l]})$ qubit pairs.
We see that the maximal distance corresponds to repeater level $11$, i.e. about $2^{11}\cdot10\mbox{ km}\approx 2\cdot10^4\mbox{ km}$ where we get a fidelity of about $0.87$. This distance is halfway around the globe, but the resources required are ridiculously high (hundreds of billions), and no optimization can change this order of magnitude significantly. 
\begin{table}[ht]
\begin{tabular}{c|c|c}
rep. level&resources&max. fidelity\\ \hline
1&$15$&0.956246\\
2&$151$&0.981122\\
3&$1480$&0.983974\\
4&$1.44\cdot10^4$&0.983830\\
5&$1.40\cdot10^5$&0.983086\\
6&$1.37\cdot10^6$&0.981557\\
7&$1.36\cdot10^7$&0.978481\\
8&$1.36\cdot10^8$&0.972266\\
9&$1.42\cdot10^9$&0.959568\\
10&$1.61\cdot10^{10}$&0.932962\\
11&$2.19\cdot10^{11}$&0.873666\\
12&-&-\\
\end{tabular}
\caption{\label{levelsAndResources} Quantum repeater with standard repeater protocol and operational and memory errors included. For the parameters of errors and initial states see the text.
The first column displays the repeater level. Level $1$ corresponds to about $10$ km, and we assumed a doubling of distance with each level. The second column contains the resources, i.e. the qubit pairs, needed to reach the corresponding level. The values in the third column are the fidelities we obtain on these levels.}
\end{table}

The Innsbruck protocol, which uses entanglement pumping for the purification, will profit even more from the fact that the states used to pump are close to binary mixtures on higher levels as compared to the pumping with Werner states (worst case, see table~\ref{purificationregime}). However, the protocol saves physical resources (logarithmic scaling with distance) at the expense of polynomial temporal overhead~\cite{Du98}. This means that pairs on higher levels do not only have to wait for the classical signals that determine whether they have undergone a successful purification step, but also for all lower levels to produce a pair they can be purified with. While the temporal resources, the waiting times, scale polynomially with distance, any waiting time enters in the exponent of the decoherence map, equation~(\ref{memoryerror}), so this poses a severe restriction on the maximal distance.
 
\begin{figure}[ht]
\includegraphics[width=0.45\textwidth,clip]{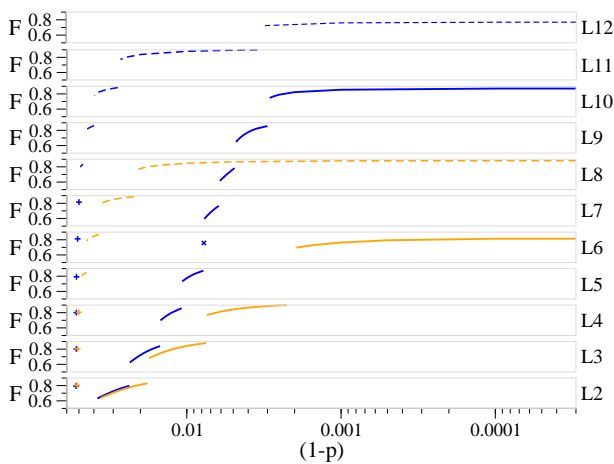}
\caption[]{\label{repeaterlevels}[color online] Maximal repeater level and fidelity $F$ as function of the operational/measurement errors ($1\!-\!p=1\!-\!\eta$). The distance on repeater level $1$ is $10$ km, every level ($L2$ to $L12$) doubles this distance. Dark lines have decoherence time $\kappa^{-1}=1$ s, light lines have $\kappa^{-1}=0.1$ s. Solid lines are a lower bound on the maximal distance for a repeater run with the Innsbruck protocol and with initial Werner states of fidelity $0.8$ on level $1$. Dashed lines show the limits of the purification map, which are an upper bound on any repeater run in error detection mode (with the exception of blind mode, section~\ref{blind}). For a more detailed discussion see text.}
\end{figure}

In Fig.~\ref{repeaterlevels} we plotted the error rates $(1-p)=(1-\eta)$ against the maximal repeater level ($L1$ to $L6$, and $L1$ to$L10$ respectively) and the maximal fidelity $F$ thereon for the Innsbruck protocol (solid curves). The upper curve (dark, solid) corresponds to a decoherence time $\kappa^{-1}=1$s, the lower to $\kappa^{-1}=0.1$s (light, solid). The initial states were Werner states of fidelity $0.8$. Before we go into details, let us examine the key features of these curves. a) On the left, we are in a regime that is dominated by the errors in operations ($1-p$) and measurements $(1-\eta)$, where we set $p=\eta$ for convenience. In this regime, a decrease in the error rate quickly leads to higher repeater levels that we can reach. b) On the right, where the errors are already small, the curve is dominated almost entirely by the decoherence time $\kappa^{-1}$. Naturally, a larger decoherence time allows for higher maximal repeater levels. In this regime we can decrease the error rates by orders of magnitude and still gain almost nothing. Note, however, that once the error rates are below $10^{-4}$ other schemes (concatenated CSS codes, quantum repeater in error correction mode) become available.

In the following we explain the details of the simulation and rules under which the plot was created.
First, we estimated the waiting times in a conservative way. The waiting time of a qubit pair on some repeater level is the time this pair has to wait either until the classical signal arrives telling us whether a purification step was successful, or until the lower levels have produced the next pair for purification (whichever takes longer). In our conservative estimate we simply add both times, that is, we wait until we get the signal, then start to build up a new pair. Decoherence affects the qubit pairs during these waiting times. With our conservative estimate we establish a lower bound on the maximally reachable distance and fidelity telling us that we can expect to reach these levels with certainty for the Innsbruck protocol. Better estimates of the waiting times will shift the solid curves upwards, but not very much: We usually gain at most $1$ level with a better estimate. When we change the initial state on the lowest level (from the Werner states with fidelity $0.8$ we used) we affect the curves only slightly. A higher fidelity for the initial Werner state (or a shape closer to a binary mixture) shifts the curves upward, and the difference becomes smaller in the region where the decoherence time dominates the plot. A lower fidelity shifts the curves downwards, and there will be a point where we lose the whole curve when we drop below the purification threshold of the first level. 
Second, for each point in the plot, we optimized the number of purification steps executed on each level. We call this the purification strategy in the following. The aim of the optimization is to reach the highest level possible. The rule when a jump from some level $l$ to a level $l+1$ occurs is the following. Assume that by some purification strategy $X$ that is optimal for level $l$ we have reached a certain fidelity $F_X^{[l]}$. Then we connect two pairs with this fidelity and get some pair with reduced fidelity $F_X^{[l+1]}$ on the next level without doing any purification on level $l+1$. If by some, usually different, purification strategy $Y$, which really does purification on level $l+1$, we can produce a level $l+1$ pair with fidelity $F_Y^{[l+1]}>F_X^{[l+1]}$, then the point in the plot moves to at least level $l+1$, where we repeat the test. If we can not find such a $Y$, then the point is drawn on level $l$ with fidelity $F_X^{[l]}$. Consider such a level-$l$-point obtained by strategy $X$. Another technical restriction is that we do not allow to execute more purification steps on level $l$ than we did on level $l-1$ in the strategy $X$. The reason is that once we can not go to a higher level, we do not have to try to save time anymore and we could in principle do infinitely many purification steps on level $l$, but this would -- while increasing the fidelity -- drastically diminish the rate with which we create pairs. Changing the above rules would alter the jumping points and fidelities, but for every reasonable restrictions the effects would not matter much. We remark that similar optimization strategies of the number of purification steps at the different repeater levels were performed by the Harvard group~\cite{TaPriv}.

The dashed lines in Fig~\ref{repeaterlevels} are the fixed points of the \textsc{dejmps}-map obtained in the way discussed in subsection~\ref{purVSreplevel}, where the dark, dashed line corresponds to $\kappa^{-1}=1$s, and the light, dashed line to $\kappa^{-1}=0.1$s. Take e.g. the point at $(1-p)=0.01$ in the upper dashed curve. There we find the value of level $11$ from table~\ref{purificationregime}. As explained in subsection~\ref{purVSreplevel}, these curves are absolute upper bounds on any repeater run in error detection mode -- with the exception of blind mode that we discuss later. Generally speaking, when we run the repeater with the standard repeater protocol, i.e. with regular entanglement purification, we will be close to the upper bound, when we run it with the Innsbruck protocol using entanglement pumping, we will be close to the lower bound. Other entanglement pumping protocols, like the Harvard protocol, can, and likely will be, even below the lower bound valid for the Innsbruck protocol. 

\begin{figure}[ht]
\includegraphics[width=0.45\textwidth,clip]{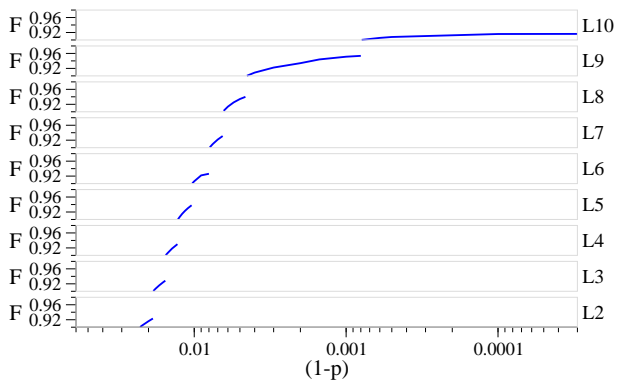}
\caption[]{\label{fidelitycriterion}[color online] Maximal repeater level and fidelity $F$ as function of the operational/measurement errors ($1\!-\!p=1\!-\!\eta$). The distance on repeater level $1$ is $10$ km, every level ($L2$ to $L10$) doubles this distance. The decoherence time is $\kappa^{-1}=1$ s, the curve is a a lower bound on the maximal distance for a repeater run with the Innsbruck protocol and with initial Werner states of fidelity $0.9$ on level $1$, showing a weak dependence on the initial fidelity as compared to Fig.~\ref{repeaterlevels}. Additionally the fidelity was required to finally be above $0.9$ on every level and all its lower levels in the repeater.}
\end{figure}

When we look at the fidelities in Fig~\ref{repeaterlevels} we see that they can be very low, and we might ask whether this is not a drawback. However, there are two things to say about this. First, even final pairs with these low fidelities can be used, e.g. for communication purposes. Under certain conditions, an eavesdropper is factored out by the purification process~\cite{As02} such that the pairs, though of low fidelity, are private. Second, we simply did not ask for pairs of higher fidelity and optimized for distance only. If, say for quantum teleportation, we need pairs of higher fidelity, we add this requirement to the rules. In Fig.~\ref{fidelitycriterion} we added the rule that on any level and on all levels below it the fidelity must finally have been above $0.9$. For the same initial conditions compared to Fig~\ref{repeaterlevels} this additional restriction would mean that the curves would move downwards. In Fig.~\ref{fidelitycriterion} we changed the initial fidelity of the Werner states to $0.9$ to comply with the new rule, so we can not assert this claim by directly comparing the two plots. However, with the changed initial fidelity we support the claim that such a change does not have a strong influence on the curves. This, we can check by comparing the plots. 

Let us sum up the key message. If we use a repeater protocol with entanglement pumping, which we do to avoid unmanageably large qubit numbers, and demand tolerable errors of one percent, then we can not reach intercontinental distances. From Fig.~\ref{repeaterlevels}, at a value of $1\!-\!p=0.01$, we read off a maximal level of $5$ for a decoherence time of $1$ s, and $3$ for $0.1$ s. If we assume better initial states and better estimates of the waiting times than our conservative ones, we might reach, say, level $7$ in the first case. But $2^7\cdot10\mbox{ km}=1280\mbox{ km}$ is still not intercontinental. There are two ways to overcome this problem: Trivially, one can try to improve the error rates or the decoherence time (see Sec.~\ref{reduceMemory}). One reaches intercontinental distances, e.g., for a decoherence time of $1$s and error rates increased by one order of magnitude, namely $0.001$. Second, one can combine protocols. On higher levels one can e.g. switch from the Innsbruck protocol to the standard repeater protocol at the expense of larger physical resources. We will come back to the question of such repeater architectures in a later section.

Note that decreasing the errors by another order of magnitude, to $10^{-4}$, does not give us much further advantage. However, at this error rate different strategies become available, and we will now turn to one of these, the repeater in error correction mode.

\subsection{Limits of the quantum repeater in error correction mode}\label{limitsECM}

In error correction mode, the repeater is limited both by the memory errors and the very stringent thresholds for operation fidelities. The first limit can be completely removed (see section~\ref{blind}) and we discuss it only shortly. The second limit remains, and we present the results for thresholds below.

\subsubsection{Limits by memory errors}
If we use purification via error correction in some repeater protocol instead of purification via error detection we still have to wait for the classical $1$-way signal to arrive in order to know which correction operation to apply. Concerning waiting times during which memory errors occur we gain nothing in this way. On the contrary, since purification ranges are much smaller than for error detection schemes~\cite{As03}, we have the following situation. We need higher fidelities in operations and measurements (at least $10^{-4}$) and are still sooner out of the game than in the error detection repeater protocols. This seems like a lose-lose situation, but we will show in section~\ref{blind} that we can overcome the problem of waiting times completely for a repeater in (concatenated) error correction mode, while this is not true for a repeater in error detection mode. For the discussion of threshold limits we will hence already assume that memory errors are absent, or, more precisely, absorbed into lowered operation fidelities.

\subsubsection{Threshold limits}\label{thresholdsECM}
Even when memory errors do not have to be taken into account explicitly, the threshold limits of operation and measurement fidelities for the whole repeater must be derived from the thresholds for entanglement purification and connection. As pointed out in section~\ref{OnewayEPP}, one can construct entanglement purification protocols from CSS codes using only one-way classical communication (i.e. the protocols run in error correction mode). Transmitting several copies of an entangled state through noisy channels and purifying them using a single step of such an entanglement purification protocol results in a single copy with increased fidelity, which can then be used to transmit quantum information via teleportation. As shown in section.~\ref{OnewayEPP}, this procedure is in fact equivalent to encoding quantum information into several qubits using this CSS code, transmitting the encoded state through the noisy channel and performing error correction (decoding) at the receiver station.

If we perform several purification steps, i.e. use output states of the previous purification round as input states for the next purification round, we can establish a similar equivalence, this time to {\em concatenated} error correction CSS codes. The number of purification steps corresponds to the number of concatenation levels of the code. This equivalence also holds when taking noise (of the form we consider here) in local control operations into account. As a consequence, entanglement purification protocols in error correction mode and quantum error correction (QEC) schemes have the same thresholds with respect to tolerable channel noise and noise in local control operations. In particular, thresholds for tolerable noise in local control operations for QEC have been estimated to be of the order of $10^{-4}$, leading to the same threshold for the corresponding one-way entanglement purification protocols. This number has to be compared to a tolerable noise of the order of several percent for entanglement purification protocols with two-way classical communication, i.e. run in error detection mode. Aschauer~\cite{As03} explicitly investigated the performance of entanglement purification protocols constructed from specific CSS codes in the presence of noisy local control operations for a simplified error model. He finds that the threshold for noise in local control operations (in his error model) is almost ten percent when using two-way classical communication, while it is of the order of 0.5 percent for one-way purification protocols. Also the tolerable channel noise (i.e. minimal required fidelity) is significantly lower for one-way purification protocols as compared to two-way protocols. 

Notice that thresholds for entanglement purification, together with the influence of noise on the connection process, determine the maximal length of the elementary segments in the quantum repeater, and also the threshold for the total repeater protocol. This threshold is even more stringent than the threshold for entanglement purification. In particular, when using one-way entanglement purification protocols, one needs to use elementary segments with smaller distance (i.e. more repeater stations), and the threshold for the repeater protocol will be significantly more stringent (by a factor of about 20-100) as compared to thresholds for the quantum repeater based on two-way entanglement purification. We finally remark that the equivalence between entanglement purification protocols and QEC schemes based on CSS codes carries over to the whole repeater protocol, where also entanglement swapping is involved. It turns out that establishing an entangled pair using the repeater protocol, i.e. by a nested sequence of entanglement purification and entanglement swapping operations, and using the pair to teleport an unknown quantum state is in fact equivalent to transmitting the quantum state in an encoded form through the noisy channel using a specific concatenated CSS code. Strictly speaking, this equivalence only holds for noise channels which are diagonal in the Pauli basis, however this is exactly the noise model we consider here.
The essential property one uses is that coding and decoding operations for CSS codes, and hence also all involved operations in the entanglement purification protocol, are Clifford operations. It follows that Pauli operators can be commuted through the coding and decoding operations as well as through the noise maps (if they are Pauli diagonal) and simply become a different Pauli operation corresponding to a (correctable) basis change. These Pauli operations appear either due to different outcomes in Bell measurements of the connection process, or due to required correction operations after establishing the error syndrome in a certain purification step. 
The communication scheme that is equivalent to the quantum repeater corresponds to using a concatenated CSS code. Concatenation comes, on the one hand, from several purification steps performed at a fixed repeater level, and, on the other hand, from the concatenated scheme of the quantum repeater to establish entangled pairs over larger and larger distances. The latter concatenation translates to a specific way in which error correction is performed at different repeater stations. At certain repeater stations, e.g. at the final station error correction at all nesting levels is performed, while at intermediate repeater stations error correction is done only up to a fixed concatenation level. For instance, at the second repeater station, only  error correction at the lowest concatenation level is executed, while at the middle repeater station (at half the distance) error correction is applied up to the second highest concatenation level. 

\section{Reducing memory errors}\label{reduceMemory}

As we have seen in the previous section, memory errors limit the possible communication distance when using a quantum repeater run in standard mode. The actual achievable distance crucially depends on the quality of local memory, characterized by the coherence time, as is evident e.g. from Fig. \ref{repeaterlevels}. If one aims to achieve quantum communication over some fixed distance, say intercontinental distance, then it is sufficient to ensure that quantum memories of sufficiently high quality are available. There are various strategies known to increase coherence times, including quantum systems with extremely weak coupling to the environment, decoherence free subspaces~\cite{DFS}, dynamical decoherence free subspaces~\cite{Ta05}, or topologically protected quantum memory~\cite{TPM}. Some experimental proposals for a quantum repeater take these strategies into account~\cite{Exp,Exp2,Kl06}, where e.g. a quantum repeater with qubits in a decoherence free subspace has been proposed in Ref.~\cite{Kl06}. Coherence times of up to $20$~s have been demonstrated experimentally~\cite{20s} for qubits in decoherence free subspaces. Although coherence times are long in this case and might be sufficient for practical purposes, they are not infinitely long, which would be required for communication over arbitrary distance. Further reduction of memory errors may be possible, at the price of increased complexity and eventually reduced error thresholds of the repeater protocol.

The complete elimination of the influence of memory errors seems only possible when using strategies from fault tolerant quantum error correction, where concatenated error correction codes are used to obtain a perfect quantum memory \cite{Steane}, leading to error threshold estimates of the order of $10^{-3}$. Notice that the problem of storage of quantum information is less demanding than the problem of processing (encoded) quantum information as it is required in fault tolerant quantum computation. When using concatenated CSS codes, only Clifford operations are required for storage, and thus one might expect less stringent error thresholds. 
The whole repeater protocol as such can still be applied in the standard fashion, and the distance between repeater stations is the same as in the case where memory errors are disregarded. This distance is essentially given by the minimal required fidelity of the two-way entanglement purification protocol. Clearly the thresholds on noisy local control operations for the whole repeater scheme are now determined by the more stringent thresholds for quantum memory. However, not at all repeater levels perfect quantum memory is required. At lower repeater levels, no quantum memory is needed. At higher repeater levels, the required storage time (and hence the required coherence time) gets larger, and high fidelity quantum memory is needed, where the effort to produce the required fidelity increases with the repeater levels. The complexity of the protection mechanism also increases, and so does the requirement on the fidelity of local control operations. Finally, at a certain repeater level, concatenated error correction codes need to be used that provide {\em perfect} quantum memory, and threshold results for such schemes can then apply.   

When concatenated error correction codes are used for local memory, it is important to note that the repeater protocol based on two-way entanglement purification (error detection mode) is still {\em inequivalent} to sending encoded quantum information through a noisy quantum channel by using again some concatenated code. For instance, the repeater stations need to be much closer in the latter case, leading to a significant overhead and possibly also to more stringent thresholds.

\section{Quantum repeater in blind mode}\label{blind}
In this section we consider a blind operational mode for the quantum repeater to overcome or lessen the limitations due to memory errors. Blind operation of the quantum repeater works for both error detection mode as well as error correction mode. In the first case, blind mode can add some additional repeater levels on top of the ones possible otherwise with reasonable overhead, in the second case it enables the quantum repeater to create entanglement over arbitrary distances, albeit with lower thresholds.

\subsection{Blind error detection mode} 

We show that the \textsc{dejmps}-protocol can be executed blindly \cite{Ca05}, i.e. without waiting for classical communication, at the price of an exponentially decreasing success probability. Entanglement swapping can also be performed blindly such that the whole repeater can run in blind mode, at least on a few levels where the additional resources, which are required to counteract the exponentially decreasing success probability, stay reasonably low.
\subsubsection{Blind purification}

Blind $2$-way purification is a variant of the standard entanglement purification in error detection mode. The only difference is that one does not wait for any classical signal to arrive, which would tell whether a purification step was successful, and thus eventually operates on ``bad'' pairs. In fact, any basis shift of input states only leads to (i) a re-interpretation of what is called a successful purification step and (ii) a new basis shift of the resulting density matrix. In this sense, the basis shifts do not matter, and the same sequence of operations (i.e. the same protocol) can be used, regardless of the initial basis shifts.

This is most evident in Eq.~(\ref{Purymap}), where entanglement purification with perfect local control operations is described. It is straightforward to see that also for noisy local operations (of the form we consider here), these properties are kept, Eq.~(\ref{imperfectPurymap}), because basis shifts (corresponding to $\sigma_z$-operations) can be commuted through noise maps that are diagonal in the Pauli basis.

This implies that, in principle, several purification steps can be performed without knowing the required correction operations. Only the interpretation of the obtained measurement outcome, and hence the decision whether the purification step was  successful or not, requires knowledge of basis shifts, and hence classical communication. Clearly, if several purification steps are performed blindly in such a way, the resulting pair is only useful if it turns out that in fact all steps correspond to successful purification steps. The success probability for the total procedure thus goes down exponentially with the number of purification steps. If the operations were perfect, the success probabilities would converge to one since also the fidelity converges to one, and the total success probability need not necessarily go down exponentially. With errors in the operations/measurements, on the other hand, the maximum reachable fidelity and thus the maximum success probability for a purification step is bounded away from one, and exponential decay of the total success probability follows.

\subsubsection{Blind swapping}
The maps for connection (entanglement swapping) do not require any specific form of the input states. Also imperfect connection processes can be performed on two pairs with arbitrary basis shifts, leading to a new pair with a new basis shift depending on measurement outcomes and the initial basis shifts. Again, this is evident from the description of the connection process when local operations are perfect (see Eq.~\ref{connectionmap}). The property is kept for noisy operations if the noise is Pauli-diagonal, Eq.~(\ref{imperfectconnectionmap}), since then we are again dealing with Clifford operations only.

\subsubsection{Blind repeater protocol}
Since both entanglement purification and swapping can be done blindly in the $2$-way, error detecting scenario the whole repeater can be operated in blind mode. Operating the repeater blindly, one can sidestep the problem of memory errors due to the long waiting times for classical signals. A new limit is set by the gate operation time, which, for entanglement pumping, still accumulates. While in principle the new maximal distance is infinite when operating the repeater with standard entanglement purification where all pairs are available in parallel, and very large for the protocols based on entanglement pumping, the success probability of the whole repeater goes down exponentially with distance. Consider the following example. 
We assume that three purification steps at each repeater level are required, $M=3$, and consider the scaling of the required resources when operating $m$ repeater levels blindly. We also assume that only two pairs are connected before re-purification. This leads to an increase of the distance by a factor of $2^m$. For simplicity we say that each purification step succeeds with a certain fixed success probability $p_{\rm suc}$ (the success probability depends on the fidelity of the initial pairs and hence is strictly speaking different for different purification steps; however, we neglect this effect since the overall scaling behavior will not be affected by this simplifying assumption). In this case, the total success probability that all involved purification processes up to repeater level $m$ were successful is given by 
\begin{equation*}
p_{\rm tot}=p_{\rm suc}^{(2^{m-1}M^m)},
\end{equation*}
and thus on average $1/p_{\rm tot}$ copies of the whole set-up (i.e. parallel channels) are required to obtain on average a single pair at the end of the procedure. Alternatively, one can say that the rate of the resulting pairs is decreased by a factor $p_{\rm tot}$. The following table illustrates that up to three additional repeater levels, $m=3$, lead to a reasonable overhead, while for $m>3$ the overheads explode and become completely impractical. For $m=3$, the possible communication distance is increased by a factor of 8, i.e. almost an order of magnitude.

\begin{table}[ht]
\begin{tabular}{|c|c|c|}
  \hline
   & $p_{\rm suc}=0.95$ & $p_{\rm suc}=0.9$ \\
  \hline \hline
  $m=1$ & $p_{\rm tot}^{-1}=1.17$ & $p_{\rm tot}^{-1}=1.37$  \\
   \hline
   $m=2$ & $p_{\rm tot}^{-1}=2.52$  & $p_{\rm tot}^{-1}=6.66$  \\
   \hline
   $m=3$ & $p_{\rm tot}^{-1}=254.6$  & $p_{\rm tot}^{-1}=8.7 \times 10^4$  \\
   \hline
   $m=4$ & $p_{\rm tot}^{-1}=2.7 \times 10^{14}$  & $p_{\rm tot}^{-1}=4.4 \times 10^{19}$  \\
  \hline
\end{tabular}
\caption{Table of required {\em additional} resources $p_{\rm tot}^{-1}$ when operating the quantum repeater in blind operational mode under the assumption that $M=3$ purification steps with constant success probability $p_{\rm suc}$ are required. Number of additional repeater levels is given by $m$, and the communication distance is increased by a factor of $2^m$.  }
\end{table}
\vspace{0.3cm}

We remark that when fewer purification steps $M$ at each repeater level are required, or more than only two elementary pairs can be connected before re-purification, one can increase the communication distance even further. One may even design the repeater scheme in such a way that at higher repeater levels (where blind mode is used) fewer purification steps $M$ are required. In this case in principle more additional repeater levels can be added while keeping the overhead moderate (for smaller $M$), and each additional repeater level not only allows one to double the distance but to increase it by a factor of $L$ (if $L$ elementary pairs can be connected), leading to a total gain of a factor of $L^m$. For instance, if $M=2$ and $L=3$, three repeater levels, $m=3$, yield an overhead factor of about 40 if $p_{\rm suc}=0.95$, while the communication distance is increased by a factor of $3^3=27$. Thus a gain of about an order of magnitude in distance with overhead of order $10^2$ seems possible, where in some favorable situations even higher gains can be expected. 
 
Because of the exponentially small success probability, blind mode is not a solution for the whole repeater in error detection mode. However, for practical purposes one may still use blind mode on a few of the topmost repeater levels at the cost of a reduced production rate of entangled pairs. In this sense, the parameter $m$ above corresponds to the {\em additional} repeater levels that are operated blindly, while low repeater levels are operated in the standard way. These last levels should be run in the parallel, standard repeater mode, since for protocols using entanglement pumping the classical signals will usually have arrived before a new pair is ready from lower levels, and it would be disadvantageous to operate blindly and to ignore the information available.

\subsection{Blind error correction mode} 

In this subsection we describe a possible solution to overcome the limitation of communication distance due to memory errors. This solution is due to the fact that the repeater can be unconditionally run blindly in error correction mode, i.e. there is no exponentially small success probability, when special error correcting codes, Calderbank-Shor-Steane (CSS) codes, are used.

\subsubsection{Blind purification and entanglement swapping}

Again, the key point is that the entanglement purification protocols can also be used if the initial bases of the pairs are shifted. More precisely, since the coding and decoding networks are based on CSS codes, all unitary operations applied in the purification protocol are Clifford operations. Therefore, any basis shift (described by some Pauli operation applied to the state before coding/decoding) can be commuted through the network, still leading to a (different) Pauli operation corresponding to a (different) basis shift. Only the interpretation of measurement outcomes when attempting to detect an error syndrome, and the final basis shift, may differ. In this sense, the classical information on measurement outcomes are not really required when performing the protocol, as the required operations are independent of eventual basis shifts. Only at the end of the procedure, when a final basis shift or correction operation needs to be determined, the classical signals containing all measurement outcomes are needed. That is, the purification protocol can be run blindly. 

The connection process by entanglement swapping is the same as in the $2$-way, error detecting scenario and can hence be performed blindly.

\subsubsection{Blind repeater protocol}

Since both entanglement purification by error correction and the connection process by entanglement swapping can be executed blindly the whole repeater can be run in blind mode. The main difference to the error detection mode is the following. Recall that in the error detection mode the purification process is probabilistic, and the total success probability hence goes down exponentially with the number of purification steps, whereas in error correction mode the purification is deterministic. Since entanglement swapping is also deterministic, the whole repeater can be run in blind error correction mode without restrictions. 
In particular this means that there are no true waiting times if concatenated error correction is used, where, similarly as in the standard repeater protocol, all pairs involved in the process are created in the very beginning. With true waiting times we mean times other than gate operation times because memory errors occurring during gate operations can be absorbed into a lowered gate fidelity. Hence, entangled pairs over arbitrary distances can be generated in this way. However, the limiting factors are the very stringent error thresholds (see~\ref{thresholdsECM}) and the huge number of qubits one would need.

We remark that despite the equivalence of the repeater run in (blind) error correction mode with direct transmission of quantum information using a certain concatenated CSS code, there is an advantage of the quantum repeater in a different respect. In particular, when one considers the time required to establish an entangled pair over distance $N$, the repeater scheme allows one to do this in $\log_2 N$ time steps where each time step corresponds to the time required for quantum communication over the distance of an elementary segment, $\tau_0$. Although the pair produced in this way is unknown at this stage until classical information arrives (which requires a time of order $N t_0$, where $t_0$ is the time for classical communication over one segment), it can nevertheless already be used for teleportation or for key distribution as outlined below. On the other hand, using error correction to protect transmitted quantum information corresponds to sending the information sequentially through quantum channels, leading to a communication time of $N \tau_0$. 

The difference in the communication time can be significant. Even when taking the additional classical communication into account, the repeater scheme may offer still advantages, in particular in situations where $\tau_0 > t_0$. This is already the case when transmitting photons through optical fibers and using free-space classical communication, however the effect is much more evident when considering quantum information transport e.g. by means of electron transmission. Such a repeater scheme is discussed in Ref.~\cite{Ta05}, where entanglement between distant quantum dots is generated by transporting electrons via charge control, connecting entangled pairs and re-purifying them. In this case, entanglement can be used to perform teleportation-based gates between far distant qubits, providing an important element for a scalable fault tolerant quantum computer architecture based on charge controlled quantum dots.

\subsection{Using unknown entangled pairs}
In both blind modes, error detection as well as error correction mode, the basis shift and hence the correct interpretation or the required correction operation remains unknown as long as all the measurement results from purification steps and connection processes are not known at the end node. 
Still, the entangled pairs produced in such a way can be useful, despite the lack of knowledge, which state is actually at hand. This can only be determined at a later stage after all classical signals arrive.
 
First, one may assume that memory errors are only relevant at intermediate repeater stations and other ways of protecting quantum information are available at starting and end points. Such an assumption is in some sense natural, as keeping produced entangled pairs as a resource requires a quantum memory anyway. In addition, even if (almost) perfect memories are available, technologically they might be difficult to realize and thus one may assume that at intermediate repeater stations memory errors play a role, while at end nodes memory errors can be avoided. 

Second, one may use the resulting entangled pair for teleportation of an unknown quantum state, thereby realizing high-fidelity quantum communication. However, the correction operations required in the teleportation protocol now do not only depend on the measurement outcomes in the teleportation process, but also on the basis of the used Bell pair (and hence on all intermediate measurement outcomes in the generation of the Bell pair). In this sense, a quantum memory is required again (at least at the end node), such the the teleported quantum state can be restored and further processed.

Third, one may use the resulting pair for quantum cryptography, i.e. to establish a secret key between $A$ and $B$. In this case, measurements are performed to either run a teleportation based version of a protocol such as the BB84 protocol~\cite{Be84}, six-state protocol~\cite{Bruss98}, Singapore protocol~\cite{Singapore}, or alternatively the E91 protocol~\cite{Ek91}. From now on, all information is classical, and storage of quantum information is no longer required. The additional information about the basis of the involved entangled pair (i.e. the outcomes of all measurements involved in the repeater protocol) may arrive at any later stage, and only lead to a re-interpretation of the measurement outcomes (i.e. the used measurement basis). Eventually, the yield of the key-distribution protocols is reduced since not all measurement bases can be used to establish a key, however key generation will still be possible. 

We remark that the possibility to operate the repeater in such a blind mode may also have consequences on the practical realization of such a device.
For the repeater operated in standard mode, it is usually argued that there should be flying qubits (usually photons) that are mapped on static qubits (atoms, ions, solid state devices, atomic ensembles) and vice versa. The flying qubits are used to distribute entanglement over noisy quantum channels, while static qubits are used to store and process quantum information at different repeater stations. However, as for a repeater operated in such a blind mode there is {\em no longer a need to store qubits}, the procession (i.e. error correction, measurements) might be performed right away on the flying qubits. In this way, one could avoid the (technically demanding) interfaces between flying and static qubits.  What remains is the requirement to process the qubits, i.e. to perform appropriated unitary operations for coding, decoding and measurements.

\section{Repeater architecture}\label{architecture}
While the quantum repeater in error correction mode offers a solution to achieve infinite communication distance, the stringent error thresholds and huge physical resources needed make it unfavorable for practical implementations.  

The most reasonable architecture of a quantum repeater, solely using error detection mode, could be the following. On the lowest levels, where classical signalling time is still short, one should employ a repeater protocol using entanglement pumping for purification. In this way, one saves physical resources. Which protocol to use exactly depends on the physical resources available, and one should always fully use the available resources to save time. Once one can not go further with this first protocol, one can switch to a protocol that operates on many copies in parallel, like the standard repeater protocol. In addition, techniques to reduce memory errors can be applied at higher repeater levels. Finally, when even the capabilities of that protocol and improved quantum memories are exhausted, one may change to operate the second protocol in blind mode on the topmost levels. The requirements for the physical resources become very demanding for the last two stages. 

The principal constraints are the distance over which one wants to establish an entangled pair, the physical resources available, and the parameters of the errors that will occur. Given these, the building of the quantum repeater is then an intricate engineering and optimization problem that has to deal with questions like: Which purification protocol do we use? Which working fidelity is best or how many purification steps do we perform on some repeater level? Which repeater protocol do we use and when do we switch to another? 
In theory this optimization can be very complicated since all these questions are dependent on each other, but in practice one will most likely also be limited in the ways one can optimize the working processes. 

We want to make one last remark on the re-use of qubits. In the standard repeater scheme, most qubits, when they have been measured, do nothing until the repeater has completed its cycle.
But one can immediately reuse any qubits that are no longer involved in the repeater process. Assume we add one more qubit at each repeater level, say $n$ qubits, then we can run a ``second wave" right after operations on the lowest level are performed under the same initial conditions we found before. If we add $n\!-\!1$ qubits on each repeater level, i.e. $n(n\!-\!1)$ qubits in total, then the ``first wave" will be complete when we start the $n^{th}$, since the repeater in standard mode needs $n$ time steps for completion when there are $n$ repeater levels. Then, the wave $n+1$ can use again the qubits of the first wave. In this way all qubits are used at all times, and for the price of the very demanding resources we get at least a very high bit-rate that is only limited by the gate operation time.  

\section{Summary}\label{summary}
We have studied the quantum repeater subject to memory errors. We have shown that memory errors imply that the standard operation mode of the repeater, error detection mode, can establish entangled pairs only over some maximal distance. To overcome this restriction, a direct solution is to reduce or correct memory errors by using methods to increase coherence times or a local quantum memory based on concatenated error correction codes. However, the complexity and requirements on accuracy of local control operations increase with the distance, and the error thresholds for quantum memory determine the error thresholds of the quantum repeater. Alternatively, one can run the repeater in error correction mode. We showed that this operation mode is equivalent to the protection of quantum information with concatenated quantum codes and has again unfavorable error thresholds. If one wants to benefit from the much higher thresholds of the standard mode using two-way entanglement purification and does not have the capability to correct memory errors, one has to accept some maximal distance and questions like scalability are no longer an issue (top down view). In their place are now questions about engineering and optimization. As an additional tool of practical importance, we described a new operation mode for the repeater called blind mode, which can help to push the limits for the maximal distance farther. In particular, one can increase the communication distance by an order of magnitude with only modest overhead in physical resources. With a given error model we analyzed different repeater protocols, the resources they require, and the maximal distance over which they can distribute entangled pairs. We suggested a general architecture for the quantum repeater that switches protocols according to demand. 

We finally also mention that free-space, satellite based quantum communication~\cite{Satellite} over long distances has been discussed as an alternative approach to the (ground-based) quantum repeater. At present it is not clear whether technological difficulties can be overcome in this proposed scheme. Notice, however, that elements of the quantum repeater and the new schemes discussed here may be adopted to enhance satellite-based schemes as well. 
Very recently, the problem of memory errors in a quantum relay~\cite{qrelay} has been addressed in Ref.~\cite{Co06}, where it was shown how to use multiplexing to increase the yield. However, this investigation does not solve the problem of memory errors in the full quantum repeater as discussed here. 
To summarize, while intercontinental quantum communication with entangled pairs, created by the quantum repeater, seems to be out of reach today, the perspective that this goal can be realized in the foreseeable future is still very promising.\\

\section*{Acknowledgements}
This work was supported by the Austrian Science Foundation (FWF) and the European Union (OLAQUI, SCALA, QICS). W.D. acknowledges support from the \"OAW through project APART, and B.K. from the FWF through project Elise-Richter.




\begin{thebibliography}{99}
\bibitem{Dur_noteCNOT}
The CNOT operation is defined by $|i\rangle_A|j\rangle_B \rightarrow |i\rangle_A|i\oplus
j\rangle_B$, where $\oplus$ denotes addition modulo 2.
\bibitem{footnote}
Strictly speaking this is not a point but a manifold in $
\mathbb{R}^4$ since whether the map can still purify a certain state depends on all $4$ coefficients $\lambda_{00},...,
\lambda_{11}$.
\end{thebibliography}
\end{document}